\documentclass[sigconf]{acmart}

\usepackage{amsmath}
\usepackage{graphicx}
\usepackage{textcomp}
\usepackage{xcolor}
\usepackage{pifont}

\usepackage[noend]{algpseudocode} 
\usepackage{mathtools}
\usepackage{leftidx}
\usepackage{booktabs} 
\usepackage{multirow}
\usepackage{makecell}

\usepackage{subfigure}
\usepackage{algorithm}
\usepackage{listings}
\usepackage{url}
\usepackage{comment}
\usepackage{natbib}[numbers]

\def\BibTeX{{\rm B\kern-.05em{\sc i\kern-.025em b}\kern-.08emT\kern-.1667em\lower.7ex\hbox{E}\kern-.125emX}}

\renewcommand\footnotetextcopyrightpermission[1]{} 
\begin{document}
\title{INSPIRIT: Optimizing Heterogeneous Task Scheduling through Adaptive Priority in Task-based Runtime Systems}
%
\author{Yiqing Wang}
\orcid{0009-0007-8809-824X}
\affiliation{
	 \institution{Beihang University}
}
\author{Xiaoyan Liu}
\affiliation{%
	\institution{Beihang University}
}
\author{Hailong Yang}
\authornote{Corresponding author}
\affiliation{%
	\institution{Beihang University}
}
\author{Xinyu Yang}
\affiliation{%
	\institution{Beihang University}
}
\author{Pengbo Wang}
\affiliation{%
	\institution{Beihang University}
}
\author{Yi Liu}
\affiliation{%
	\institution{Beihang University}
}
\author{Zhongzhi Luan}
\affiliation{%
	\institution{Beihang University}
}\author{Depei Qian}
\affiliation{%
	\institution{Beihang University}
}

%

%

\begin{abstract}
As modern HPC computing platforms become increasingly heterogeneous, it is challenging for programmers to fully leverage the computation power of massive parallelism offered by such heterogeneity. Consequently, task-based runtime systems have been proposed as an intermediate layer to hide the complex heterogeneity from the application programmers. The core functionality of these systems is to realize efficient task-to-resource mapping in the form of Directed Acyclic Graph (DAG) scheduling. However, existing scheduling schemes face several drawbacks to determine task priorities due to the heavy reliance on domain knowledge or failure to efficiently exploit the interaction of application and hardware characteristics. In this paper, we propose INSPIRIT, an efficient and lightweight scheduling framework with adaptive priority designed for task-based runtime systems. INSPIRIT introduces two novel task attributes \textit{inspiring ability} and \textit{inspiring efficiency} for dictating scheduling, eliminating the need for application domain knowledge. In addition, INSPIRIT jointly considers runtime information such as ready tasks in worker queues to guide task scheduling. This approach exposes more performance opportunities in heterogeneous hardware at runtime while effectively reducing the overhead for adjusting task priorities. Our evaluation results demonstrate that INSPIRIT achieves superior performance compared to cutting edge scheduling schemes on both synthesized and real-world task DAGs.


\end{abstract}

\maketitle

\renewcommand{\algorithmicrequire}{\textbf{Input:}}
\renewcommand{\algorithmicensure}{\textbf{Output:}}

\newcommand{\tabincell}[2]{\begin{tabular}[t]{@{}#1@{}}#2\end{tabular}}

\lstset{
	basicstyle          =   \footnotesize\sffamily,          
	keywordstyle        =   \bfseries,          
	commentstyle        =   \rmfamily\itshape,  
	stringstyle         =   \ttfamily,  
	numbers             =   left,   
	showspaces          =   false,  
	numberstyle         =   \ttfamily,    
	showstringspaces    =   false,
	captionpos          =   t,      
	frame               =   lrtb,   
	breaklines, 
	columns=flexible, 
}





\newcommand{\red}[1] {\textcolor{red}{\it{#1}}}
\newcommand{\revision}[1] {\textcolor{blue}{{#1}}}
\newcommand{\secondrevision}[1] {\textcolor{black}{{#1}}}
\newcommand{\shepherd}[1] {\textcolor{black}{{#1}}}
\newcommand{\finalshepherd}[1] {\textcolor{black}{{#1}}}
\newcommand{\remove}[1] {{\textcolor{red}{\sout{#1}}}}
\newcommand{\add}[1] {{\textcolor{red}{\underline{#1}}}}
\newcommand{\egg}[1] {}
\newcommand{\separate}[1] {\textbf{\center ======  #1  ====== }}
\newcommand{\smalltitle}[1] {\vspace{6pt} \noindent \textbf{#1}}
\def\approx{$\sim$}

\def\receivers{\textbf Q}
\def\capacity{{ \textbf{c}}}
\def\upperBound{{\mathcal D}}
\def\schedule{{\mathcal S}}
\def\throughput{{\cal T}}
\def\period{{\textbf Z}}

\newcommand{\prob}[1]{{\textbf{Pr}\left(#1\right)}}
\newcommand{\mcell}[2]{ \parbox[h]{#1}{ \vspace{0.5mm} #2 \vspace{0.5mm}}}



\renewcommand{\algorithmicrequire}{\textbf{Input:}}
\renewcommand{\algorithmicensure}{\textbf{Output:}}

\newcommand{\KB}{~KB}
\newcommand{\MB}{~MB}
\newcommand{\GB}{~GB}
\newcommand{\MBs}{~MB/s}
\newcommand{\mus}{~$\mu s$}
\newcommand{\ms}{~$ms$}

\newcommand{\eg}{\textit{e.g.}}
\newcommand{\ie}{\textit{i.e.}}
\newcommand{\etc}{\textit{etc.}}
\newcommand{\aka}{\textit{a.k.a.}}
\section{Introduction}
\label{sec:introduction}


Modern HPC computing platforms are evolving towards more heterogeneity for both computation capability and memory efficiency~\cite{xie2023merchandiser,atchley2023frontier,ding2023evaluating,huang2023rm}. However, the ever-increasing complexity of heterogeneity hinders programmers to effectively harness such performance opportunities. Specifically, optimizing application performance on heterogeneous platforms poses significant challenges due to the intricate interaction of application (e.g., control flow and data flow) and hardware (e.g., computation capability, data access bandwidth) characteristics. To address these challenges, several task-based runtime systems such as Legion~\cite{6468504}, KAAPI~\cite{gautier2007kaapi}, PaRSEC~\cite{bosilca2013parsec}, StarPU~\cite{augonnet2009starpu}, OpenMP~\cite{chandra2001parallel}, OmpSs~\cite{bueno2011productive} and Tascell~\cite{hiraishi2009backtracking} have been proposed to serve as an intermediate layer to hide the complex heterogeneity from the application programmers. The core functionality of these systems is to decompose computation into tasks and manipulate the task mapping to the hardwares efficiently. For applications, these systems provide a task-based programming paradigm that enables applications to manage the computation into fine-grained tasks and organize the computation dataflow into Direct Acyclic Graph (DAG) for guiding task execution at runtime. Each vertex in the DAG represents a specific task to be performed, whereas the edges represent dependencies between these tasks. For hardwares, to better manage the heterogeneous resources, these systems commonly represent the heterogeneous processing units (e.g., CPU and GPU) as workers with individual worker queues that accept tasks for execution. 




As the scheduling technique lies in the central of task-based runtime system, its efficiency dominates the effectiveness of task mapping to the hardware resources, and thus the performance of task-based applications. The existing task-based scheduling techniques can be mainly classified into two categories such as static DAG scheduling~\cite{kwok1996dynamic, topcuoglu2002performance, Shahul2010SchedulingTG, constrainBasedScheduling, driss2015new, SFXTeixeira2023AutomatedMO, thost2021directed, zhang2019d} and dynamic DAG scheduling~\cite{2013An, wu2018adaptive, blumofe1999scheduling, workStealing2}. For static scheduling, the mostly adopted techniques include heuristic search-based scheduling\cite{Shahul2010SchedulingTG, constrainBasedScheduling, driss2015new, SFXTeixeira2023AutomatedMO}, deep learning-based scheduling~\cite{thost2021directed, zhang2019d, wu2018adaptive} and list scheduling (discussed below). The drawback of heuristic search-based scheduling is that it requires iterative tuning when adapting to new hardwares or applications, incurring high time overhead. Whereas, deep learning-based scheduling depends on highly accurate model that incurs high modeling complexity, extensive training time and resource consumption. For dynamic scheduling, the proposed techniques further leverage the runtime information for task mapping in addition to the static scheduling. For example, the existing work~\cite{wu2018adaptive} applies deep reinforcement learning to achieve runtime feedback for fine-tuning the task mapping. However, the limitation still exists as the static scheduling for the high modeling complexity and training overhead.


Due to the low complexity and overhead compared to the aforementioned scheduling techniques, list scheduling such as the HEFT and CPOP~\cite{topcuoglu2002performance} has been widely adopted in existing task-based runtime systems. List scheduling efficiently transforms the task scheduling into the determining of task priorities. Therefore, task-based runtime systems can effectively manipulate task execution based on task priorities and historical runtime information. However, existing works~\cite{tlrCholeskyParsec} often determine task priorities for specific applications based on user-configured task weights derived from domain knowledge, demanding a profound understanding domain knowledge and thus hardly portable across different platforms. Therefore, when applied to a different platform, the task priorities either require another time-consuming efforts of tuning, or achieve sub-optimal performance. There are also research works~\cite{kwok1996dynamic} to determine the task priorities based on critical path analysis, however these methods are extremely time-consuming, and difficult to scale to a large number of tasks. Although deep reinforcement learning models~\cite{wu2018adaptive} have also been applied to reduce the computation complexity to determine task priorities during runtime, these models are difficult to train due to limited datasets, and thus exhibit poor generality across different application and hardware scenarios.

In sum, existing methods for optimizing the task scheduling through determining the task priorities either rely on the domain knowledge to identify the optimal priority setting based on user-configured task weights derived from domain knowledge (e.g., static scheduling) that is hardly sustainable when considering the complex interactions between application characteristics and hardware architectures, or require high computation complexity and large volume of training data (e.g., dynamic scheduling) that is infeasible to obtain optimal task priorities in a reasonable time. In order to achieve better task priorities for optimizing the scheduling performance, we argue that \textit{1)} the optimal task priorities should be determined by the high-performant task mappings that jointly consider application and hardware characteristics, and \textit{2)} the optimal task priorities should be adjusted adaptively during runtime considering both the available computation capacity and the amount of available tasks. For example, when there are more computation resource available, the runtime should schedule more parallel tasks to increase the amount of computation, whereas when there are not enough tasks available, the runtime should prioritize the tasks on the critical path to generate more tasks.

In this paper, we propose INSPIRIT, an efficient and lightweight task scheduling framework designed for task-based runtime systems on heterogeneous hardware resources. Inspired by prior research, we dynamically adjust task priorities during runtime to manipulate task scheduling for better performance, considering both runtime information and hardware characteristics. However, unlike previous approaches, INSPIRIT proposes two novel task attributes such as \textit{inspiring ability} and \textit{inspiring efficiency}, which can effectively avoid the reliance on domain knowledge, and better determine the optimal task priorities. In addition, INSPIRIT adopts the number of ready tasks (depicted as \textit{nready}) in all worker queues as an indicator to guide task execution, which can effectively reduce the overhead of task mapping adjustment different hardwares while concurrently monitoring the runtime state.

Specifically, INSPIRIT extracts task inspiring ability and inspiring efficiency by runtime execution data and user defined cost models,  which can be taken directly. Then, INSPIRIT verified that the above attributes of task, inspiring ability and inspiring efficiency, are correlated with the number of tasks that could be executed in the environment, \textit{nready}. Therefore, INSPIRIT regulates \textit{nready} in the worker queue by introducing ascending slopes, task window size and other parameters to ensure that the performance of the machine is as full as possible, thus indirectly guiding the execution of the task. By extracting task attributes and monitoring \textit{nready}, INSPIRIT can schedule efficiently on different applications and hardware with acceptable implementation overhead. In detail, INSPIRIT adaptively adjusts the priorities of tasks and changes the execution order of tasks. Priorities then in accordance with task dependencies offered by DAG determine different task execution orders. We evaluate INSPIRIT on Cholesky, LU, and several hand-generated applications to achieve performance speedup ranging from 1.03$\times$ $\sim$ 3.22$\times$.

With INSPIRIT, users can map tasks on different DAGs to different hardware resources through runtime task-based systems without specific domain knowledge and hardware knowledge and achieve impressive performance. Besides, INSPIRIT is used as a plugin for the scheduler which users does not need to be aware of. We also extend Taskbench~\cite{slaughter2020task} with an automatic task graph generation tool, which is convenient for users to generate numerous DAGs similar to real scenes to quickly test customized policies on backends of task-based runtime systems such as PaRSEC, Legion and StarPU. 

Specifically, this paper makes the following contributions:
\begin{itemize}
	\item We propose a new approach to determine tasks priorities effectively during runtime using two novel scheduling concepts, \textit{inspiring capability} and \textit{inspiring efficiency}, which can not only efficiently adapt to various applications without large tuning overhead, but also achieve better performance across different hardware platforms.
	\item We implement INSPIRIT, a task-based scheduling framework with adaptive priority to improve the performance of tasks on heterogeneous hardwares during runtime. INSPIRIT also provides better extensibility by only using abstracted collectively by existing task-based runtime systems.
	\item We evaluate INSPIRIT with both synthesized and real-world programs such as Cholesky and LU solvers to demonstrate its effectiveness. The experimental results show that INSPIRIT achieves superior performance compared to several cutting-edge task priority schemes.
\end{itemize}

The remainder of this paper is organized as follows. Section~\ref{sec:background} describes the existing task-based runtime systems, with their limitations discussed in Section~\ref{sec:motivation}. Our optimized task scheduling framework INSPIRIT is presented in Section~\ref{sec:methodology}. Section~\ref{sec:evaluation} presents the evaluation results of the INSPIRIT on both synthesized and real-world DAGs. We discuss the related work in Section~\ref{sec:relatedwork}, and conclude this paper in Section~\ref{sec:conclusion}.
\section{Background}
\label{sec:background}



\subsection{Task-based runtime system}
\label{sec:background:task-based_runtime_system}

\textbf{Task-based Parallel Paradigm - }Task-based runtime systems are designed to support the task-based parallel paradigm, offering higher versatility compared to traditional paradigms like shared memory (pthreads) or message passing (MPI). This paradigm simplifies workload mapping and concurrency control. In this approach, an application is represented as a Directed Acyclic Graph (DAG), where nodes are tasks (basic building blocks) and edges signify task dependencies. Tasks, defined as operation groups with specific data dependencies, can be executed independently once their dependencies are met, inherently enabling parallel execution as multiple tasks can run simultaneously without interference.

\textbf{Task-based Runtime System - }A task-based runtime system creates an environment for executing such programs, typically comprising four main components: dependency monitor, data manager, task scheduler, and workers. The dependency monitor oversees task submission and execution, marking a task as ready once its dependencies are resolved. The data manager orchestrates data transfers, maintaining parallelism transparency for tasks. The task scheduler allocates ready tasks to suitable workers to ensure high performance execution of the application. Workers are an abstraction for underlying parallel computing devices, and each worker is capable to execute certain kinds of tasks.

\textbf{StarPU - }StarPU exemplifies a task-based runtime system for hybrid architectures. It provides optimized infrastructure like task dependency management, heterogeneous scheduling, data transfer, and communication. It has been utilized in various high performance applications such as HiCMA~\cite{ltaief2016hicma} and PaStiX~\cite{henon2002pastix}. In a starpu program, tasks are defined by a \textit{codelet}, data handles, data access mode, and extra dependency infomation. A \textit{codelet} is a template for similar tasks with varying inputs, comprising essential task information and multiple function pointers for different hardware-specific implementations. Task dependencies in StarPU are either explicitly stated or inferred from data handles and their access modes, aided by StarPU's virtual memory system.

In StarPU applications, users submit tasks by defining codelets, data handles, and access modes. The dependency manager then adds these tasks to a monitoring pool. When a task's dependencies are fulfilled, it is queued for scheduler to dispatch. The scheduler distributes those ready tasks to workers for execution using customizable scheduling algorithm and instructs the data manager to start data transfer (which we shall call \textit{push}). Workers maintain individual task queues, and pull task from the queue for executing as they become available (which we shall call \textit{pop}). The dependency manager releases all dependencies with respect to a task when it is completed and concludes the program upon completion of all tasks.

\textbf{Scheduling Policies in StarPU - }As task-based approach relieves manual labor of parallel programming, the responsibility for ensuring the performance of parallel programs has therefore shifted to the runtime system, which has elevated the importance of task scheduling. Task scheduling on multiprocessors is a NP-Complete problem that has long been an important subject of computer science. StarPU incorporates advanced deque model scheduling algorithms alongside traditional methods like random and work-stealing. The deque model based schedulers utilize worker-specific queues, with the scheduler distributing tasks to them from the scheduler's own queue. In the basic Deque Model (DM) scheduler, tasks are scheduled in the sequence they become available, with consideration given to task performance models to minimize overall task termination times. The Deque Model Data Aware (DMDA) algorithm extends this approach by factoring in data transfer times, demonstrating a notable success. However, this model does not account for the varying scheduling needs of different tasks due to the lack of domain-specific knowledge in the scheduler. To address this, priorities are assigned to tasks to facilitate more effective scheduling, a feature implemented in the StarPU's Deque Model Data Aware Priority (DMDAP) scheduler.

\subsection{Illustrating Priority in Scheduling}
\label{sec:background:priority}
\begin{figure}[tb]
	\centering
	\includegraphics[width=0.8\linewidth]{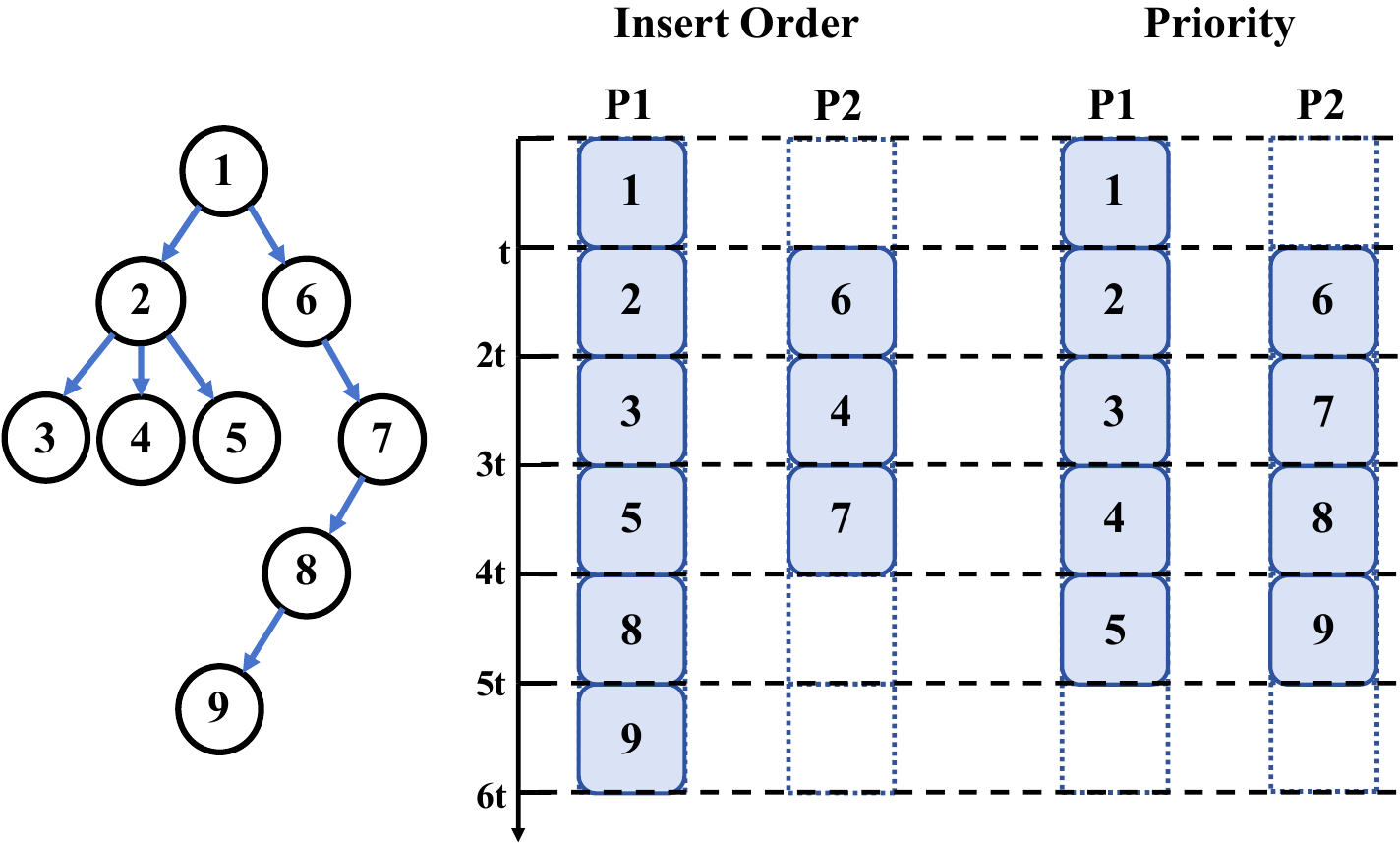}
	\caption{Demonstrating the Effectiveness of Priority in Scheduling.}
	\label{fig:why_p_works}
\end{figure}
Priority-aware schedulers leverage assigned priorities to enhance scheduling decisions. To illustrate how priorities boost application performance, consider the example task DAG in Figure~\ref{fig:why_p_works}. Assuming the program runs on a homogeneous architecture with 2 processors, and each homogeneous task takes the same time \textit{t} to complete. Without priorities, tasks are scheduled in the order they are submitted under the DMDA scheduler. The scheduling decision remains the same when there are no more ready tasks than the number of processors. At \textit{2t}, the number of tasks exceeds the number of processors. The DMDA scheduler executes tasks in their insertion order, and processors are not fully utilized after \textit{4t}. By setting appropriate priorities, the scheduler adjusts task execution order, ensuring that all processors remain fully utilized after \textit{4t}.
\section{Motivation}
\label{sec:motivation}
Our work is motivated by three key observations regarding the characteristics of DAG scheduling based on priority-based strategies through the following experiments. Regarding applications, we employ tiled Cholesky factorization as a motivating example. This approach assumes that Cholesky tasks receive data blocks of size 960 $\times$ 960 $\times$ 4 Bytes. We vary the matrix dimensions processed by Cholesky (ensuring that the number of elements along each edge is a multiple of 960) to mimic variations in application scale. On the hardware front, we evaluate diverse heterogeneous hardware environments, with specifics delineated in Sections~\ref{sec:motivation:observation1} and ~\ref{sec:motivation:observation2}. Additionally, to mitigate the impact of runtime variability, we conduct each experiment ten times. We discard the initial three results and compute the average of the remaining runs to determine the final Gflops for task execution. Moreover, based on our observations of the experimental results, we further propose \textit{inspiring ability} and \textit{inspiring efficiency} to guide scheduling, as detailed in Section~\ref{sec:motivation:observation3}.


\begin{figure}[htbp]
    \centering
    \subfigure[Performance on Cholesky under DMDAP vs DMDA policies.]{%
        \includegraphics[width=0.3\textwidth]{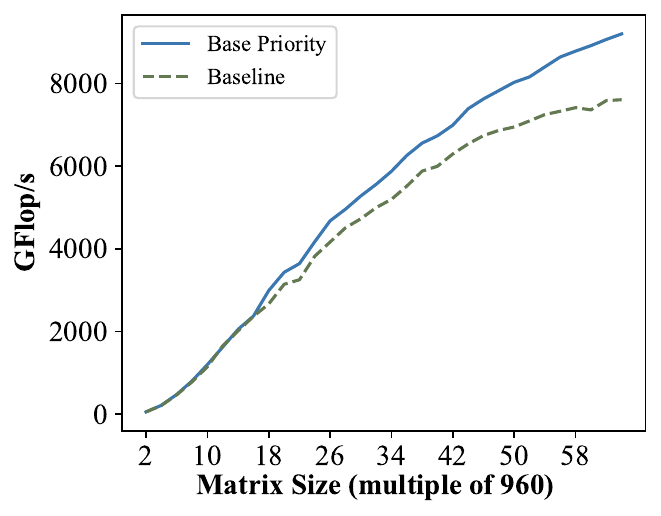}%
        \label{subfig:1_dmdap_btr_dmda}%
    }
    \subfigure[Relative speedup of Cholesky on different applications and hardware under different DMDAP policies.]{%
        \includegraphics[width=0.5\textwidth]{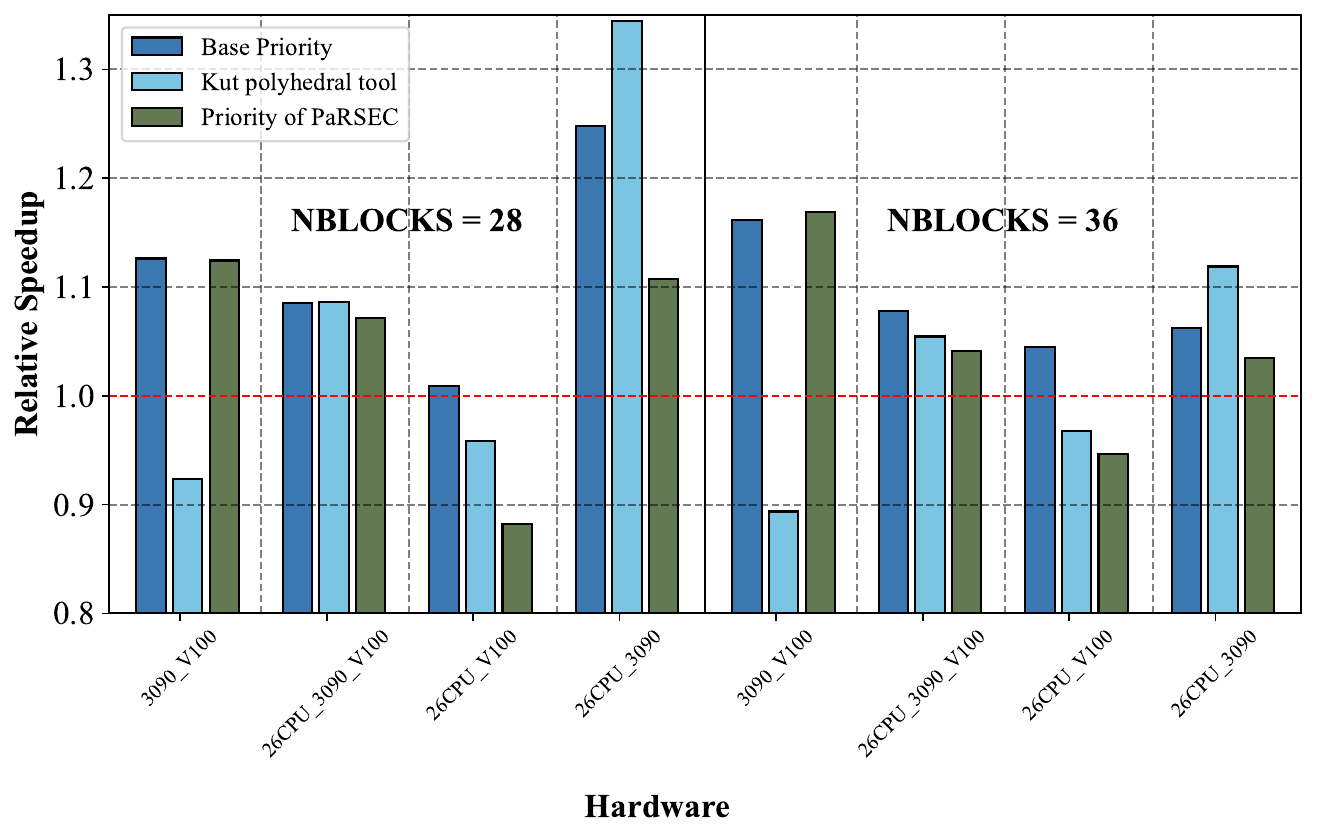}%
        \label{subfig:2_dif_env_dif_p}%
    }
    \subfigure[The Distinction in Task Execution Between DMDA and INSPIRIT at the Same Time (Blue for Executed Tasks; Red for Waiting Tasks).]{%
        \includegraphics[width=0.24\textwidth]{figures/3_missing_oppo.pdf}%
        \hspace{5pt}
        \includegraphics[width=0.24\textwidth]{figures/3_missing_oppo_2.pdf}%
        \label{subfig:3_missing_oppo}%
    }
    \caption{Observations on Tiled Cholesky Factorization.}
    \label{fig:four_images}
\end{figure}

\subsection{Observation 1: Dmdap with good priorities has impressive performance}
\label{sec:motivation:observation1} 
In our experiments, we evaluated the throughput variations of the Cholesky decomposition at various scales using both DMDA and DMDAP, as depicted in Figure~\ref{subfig:1_dmdap_btr_dmda}. The results demonstrate a consistently superior performance of DMDAP over DMDA given good priorities. This indicates that priority can directly impact the performance of task scheduling. Additionally, it can be observed from this figure that without altering the type of task, but merely increasing its scale, the effectiveness of the prioritization used in Cholesky correspondingly escalates. This observation highlights the impressive performance of carefully chosen priority.

\subsection{Observation 2: Priorities strongly dependent on the application and hardware}
\label{sec:motivation:observation2} 
Subsequently, we evaluated the throughput of the Cholesky decomposition across different scales, diverse hardware environments and different scheduling strategies (i.e. DMDA and DMDAP with various priority settings), as illustrated in Figure~\ref{subfig:2_dif_env_dif_p}. Even within identical hardware contexts, the optimal priority shifts in response to changes in data scale. For instance, in the V100\_3090 hardware environment, the best priorities for NBLOCKS set at 28 and 36 are 'Base Priority' and 'Priority of PaRSEC', respectively. This demonstrates that the efficacy of prioritization is heavily dependent on the application at hand. Concurrently, it is noticeable that as hardware specifications change, the optimal priorities are subject to alter accordingly. Take Cholesky of NBLOCKS 36 as an example, for 3 different hardware configurations (3090\_V100, 26CPU\_V100\_3090, 26CPU\_3090), the optimal priorities are identified as 'Priority of PaRSEC', 'Base Priority' and 'Priority by Kut's polyhedral tool', respectively. These findings underscore the profound influence of hardware resources on the efficacy of priority settings.

\subsection{Observation 3: \textit{Inspiring ability} and \textit{inspiring efficiency} can generate impressive priorities}
\label{sec:motivation:observation3} 

Although priority-based scheduling effectively reduces the complexity of the task scheduling, the generation of priorities remains a complex issue, since the performance gain from adopting a specific priority is highly dependent on both the application and hardware.In light of this, we propose a novel scheduling strategy based on the observation of task execution counts within a unit time window.The central idea of this strategy is to leverage the load-balancing capability of DMDA. In identical time boundaries, the more tasks DMDA can perceive, the more equitably it can achieve load balancing, accordingly better performance. Specifically, there exists an upper limit, $n_{bound}$, to the number of tasks a machine can execute within a given time period. This $n_{bound}$ is closely related to the hardware configurations and the type of tasks executed. If the number of tasks executed within a unit time window is less than $n_{bound}$, it suggests a possible shortage of ready tasks for schedule. To address this, we abstract two task attributes: \textit{inspiring ability} and \textit{inspiring efficiency}, which respectively represent the number of dependencies released upon task completion and the number of dependencies a task can release per unit time. If there are not enough tasks scheduled in the environment, tasks with higher \textit{inspiring efficiency} should be executed immediately. Otherwise, if there has already been enough tasks for schedule, tasks with higher \textit{inspiring ability} should be executed for more tasks to be available in the future.

Figure\ref{subfig:3_missing_oppo} illustrates a segment of the DAG for Cholesky factorization, a prevalent matrix decomposition method in numerical algorithms. The conventional DMDA scheduler yields the scheduling outcome depicted in the left section of Figure\ref{subfig:3_missing_oppo}, where tasks in a certain level are only executed after all tasks in the preceding level is scheduled. This sequencing can potentially impede parallelism in the final stages of execution due to fewer tasks in a level. INSPIRIT, by integrating concepts of \textit{inspiring ability} and \textit{inspiring efficiency}, dynamically assigns priorities to tasks, resulting in an optimized schedule. At the commencement of execution, when parallelism is plentiful, INSPIRIT prioritizes tasks with higher \textit{inspiring ability} to foster additional potential tasks, thereby sustaining parallelism over an extended duration. As demonstrated in the right section of Figure\ref{subfig:3_missing_oppo}, INSPIRIT accelerates the commencement of subsequent level tasks, temporarily reducing parallelism but facilitating future parallelism. Towards the execution's conclusion, when parallelism dwindles, INSPIRIT shifts focus to \textit{inspiring efficiency} to ensure maximal hardware utilization.


\begin{figure}[tb]
	\centering
	\includegraphics[width=\linewidth]{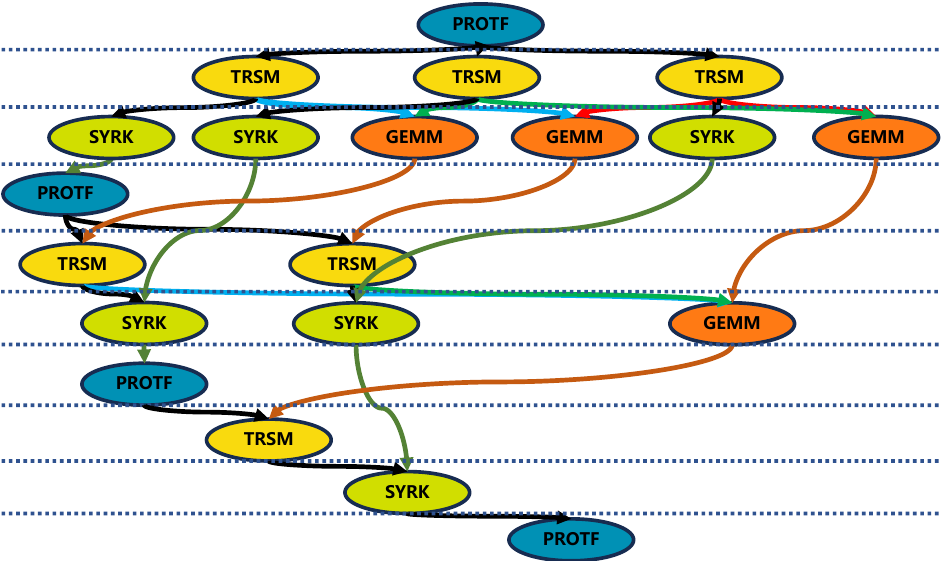}
	\caption{Tile Cholesky factorization DAG ($4 \times 4 \times 960 \times 960$).} 
	\label{fig:Cholesky_dag}
\end{figure}


\section{Methodology and Implementation}
\label{sec:methodology}

\subsection{Design Overview}
\label{sec:methodology:overview}

\begin{figure}[tb]
	\centering
	\includegraphics[width=1\linewidth]{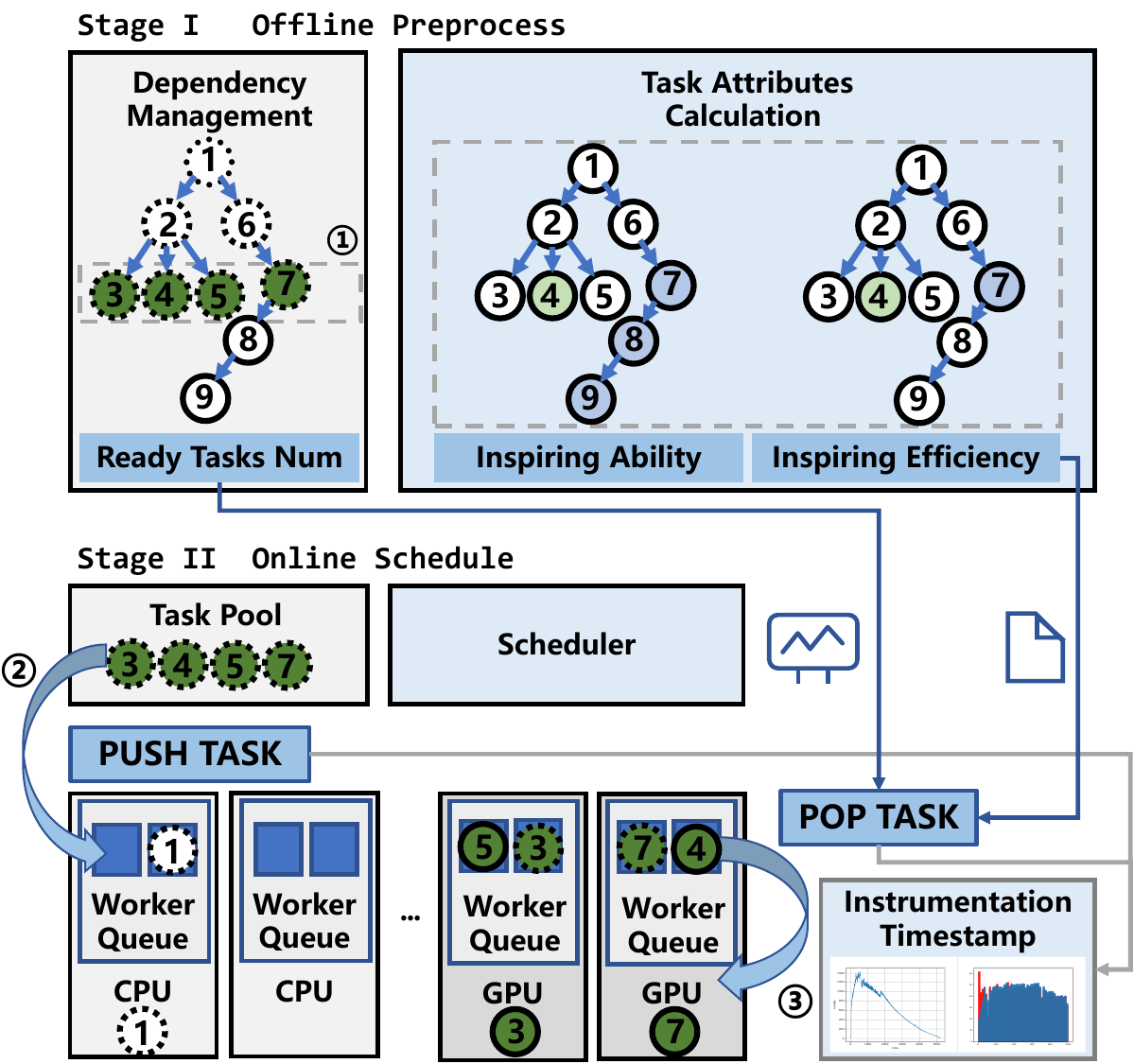}
	\caption{The design overview of INSPIRIT.}
	\label{fig:overview}
\end{figure}

As shown in Figure~\ref{fig:overview}, we implement the scheduling framework, INSPIRIT. Initially, INSPIRIT constructs the DAG and calculates task attributes offline. Subsequently, INSPIRIT regulates the number of ready tasks (\texttt{Nready}) by assigning priority to tasks based on their attributes. Using the two attributes, INSPIRIT assigns higher priorities to (1) high \textit{inspiring efficiency} tasks when insufficient numbers of ready tasks exist (i.e., executed tasks fall behind machine's capacity.), and (2) high \textit{inspiring ability} tasks when sufficient numbers of ready tasks exist. For instance, considering the task flow in Figure~\ref{fig:overview}, as illustrated in \ding{182}, ready tasks that have satisfied both task dependencies and data dependencies are scheduled to a task pool. The scheduler, as depicted in \ding{183}, according to the specific scheduling policy, dispatches tasks from the task pool to the processor's worker queue for execution. By default, each processor employs a First-In-First-Out(FIFO) strategy to execute the head task in the queue initially. For instance, task 3 is processed by the GPU worker before task 5. However, task 7 exhibits a higher \textit{inspiring ability}. Consequently, under the guidance of task attributes, as shown in \ding{184}, task 7 takes precedence over task 4 in execution, thereby resolving the dependency on task 8 and enhancing the overall application performance. A detailed explanation for the performance improvement has been shown with a simplified example in Section~\ref{sec:background:priority}.

The implementation of INSPIRIT entails three key steps: task attribute definition, task attribute computation, and \texttt{Nready} regulation.
To begin, we introduce definitions of two fundamental task attributes: \textit{inspiring ability} and \textit{inspiring efficiency}. We subsequently validate that prioritizing tasks according to these two metrics to adjust task execution order can influence the number of ready tasks, as expounded in the priority definition section (Section~\ref{sec:methodology:priority_definition}). Next, we elucidate the methodologies for computing \textit{inspiring ability} and \textit{inspiring efficiency}, as detailed in the priority computation section (Section~\ref{sec:methodology:priority_computation}). Subsequently, we regulate \texttt{Nready} to ensure consistently high hardware utilization by adjusting execution order according to the aforementioned metrics and key parameters (slope and time window size), as detailed in the \texttt{Nready} regulation section (Section~\ref{sec:methodology:regulating_nready}). Lastly, we provide detailed implementation for integrating INSPIRIT into existing task-based runtime systems, as covered in the implementation details section (Section~\ref{sec:methodology:implementation_details}), including Taskbench with auto graphs, instrumentation timestamp module.

\subsection{Priority Definition: A Foundation for Task Scheduling}
\label{sec:methodology:priority_definition}
In this section, we delve into the abstraction and definition of task attributes within the INSPIRIT, emphasizing the pivotal role of \textit{inspiring ability} and \textit{inspiring efficiency} in priority generation and \texttt{Nready} regulation. Drawing inspiration from the DMDAP strategy, INSPIRIT assigns task priorities based on their inherent attributes to orchestrate the execution sequence effectively. We initially elaborate on the core tenets of INSPIRIT and its differentiation from critical-path based schedulers.
Both critical-path-based schedulers and INSPIRIT aim to accelerate programs, with critical-path-based schedulers focusing on minimizing execution time by computing task-to-hardware mappings and execution sequences, albeit an NP-hard problem. In contrast, INSPIRIT maximizes hardware utilization by ensuring sufficient ready tasks. 
Hence, for the continual sufficient ready tasks within the task pool, tasks designated with high priority must fulfill one of two criteria: either they must swiftly release a significant number of dependencies, thereby guaranteeing consistent long-term machine throughput, or they must promptly release a comparatively large number of dependencies, thus enabling an immediate enhancement in short-term machine throughput.

\begin{figure}[tb]
	\centering
	\includegraphics[width=0.25\linewidth]{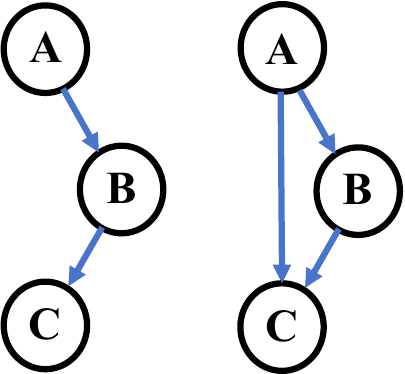}
	\caption{A description of \textit{inspiring ability}.}
	\label{fig:inspire_ability}
\end{figure}

In accordance with the outlined criteria for priority assignment, INSPIRIT introduces two pivotal metrics as task attributes: \textit{inspiring ability} and \textit{inspiring efficiency}. \textit{Inspiring ability} measures the number of resolved dependencies upon a task completion. As illustrated in Figure~\ref{fig:inspire_ability}, while node B directly relies on node A, the execution of node C is contingent upon node A's completion. Consequently, the execution of node A effectively liberates two dependencies: one to node B and another to node C. Conversely, \textit{inspiring efficiency} denotes the number of resolved dependencies per unit time of a task. The unit time is dynamically tailored based on the application, ensuring a diverse range of inspiring efficiencies among nodes (if set too low, the number of executable successor nodes per task tends towards zero; if set too high, inspiring efficiency converges towards \textit{inspiring ability}).
With the concepts of \textit{inspiring ability} and \textit{inspiring efficiency}, we can infer that prioritizing tasks with higher \textit{inspiring ability} will help release more dependencies within the DAG, whereas prioritizing tasks with better data locality and higher \textit{inspiring efficiency} can significantly enhance machine throughput in the short term.

\subsection{Priority Computation: Algorithmic Insights and Procedures}
\label{sec:methodology:priority_computation}
In this section, we delve into the algorithmic insights and procedures essential for the computation of task priorities. 
INSPIRIT computes both \textit{inspiring ability} and \textit{inspiring efficiency} by utilizing the interfaces provided by existing task-based runtime systems offline. To calculate \textit{inspiring ability}, we require application DAGs generated by the dependency management module of task-based runtime systems. The \textit{inspiring ability} of each task can be derived by traversing the entire graph, starting with nodes having zero in-degrees, and recording the number of subsequent tasks for each. For example, in a fully connected graph with only one task having zero in-degree, the value of \textit{inspiring ability} for the node with zero in-degree equals the total number of nodes in the graph. This approach maintains an acceptable computational complexity because it involves examining all tasks in the graph once, introducing necessary global information into the scheduling process. 

To compute \textit{inspiring efficiency}, we utilize the performance profiling module of task-based runtime systems to gather the average execution times for each type of task node on heterogeneous hardware offline. We choose to use the execution time on GPUs as the task execution time, prioritizing the full utilization of GPU computing power despite differences in execution times between CPU and GPU nodes. Next, we establish a unit time and calculate the number of subsequent tasks that each task can complete within the specified time as \textit{inspiring efficiency}. We test whether the \textit{inspiring efficiency} values corresponding to this unit time can distinguish between nodes of different topological layers and task types. If not, we iteratively adjust the unit time based on the \textit{inspiring efficiency} values until it effectively distinguishes between nodes of different layers and types. Therefore, the complexity of \textit{inspiring efficiency} calculation remains acceptable, as each node only needs to inspect its successor nodes executed within the unit time, even if multiple iterations are required during the computation process.

\subsection{Regulating \texttt{Nready}: Adjust \texttt{Nready} with Adaptive Priority}
\label{sec:methodology:regulating_nready}
The fundamental idea behind INSPIRIT ensuring sustained high machine performance across different hardware and applications is by regulating the fluctuations of \texttt{Nready}. \texttt{Nready} represents the number of ready tasks in the task pool that have satisfied task and data dependencies. The value of \texttt{Nready} increases with the resolve of new dependent tasks, yet decreases as the number of tasks executed by the machine grows.

INSPIRIT regulates \texttt{Nready} based on two direct principles: during periods of \texttt{Nready} increase or decrease, the value of \texttt{Nready} should undergo minimal abrupt changes; during stable periods of \texttt{Nready}, the value of \texttt{Nready} should ideally be greater than or equal to the number of hardware threads.
To implement these principles, INSPIRIT assigns priorities in three methods under different circumstances: prioritizing \textit{inspiring ability}, prioritizing \textit{inspiring efficiency}, and prioritizing data locality. 
Assuming the program prioritizes tasks with high \textit{inspiring ability}, it may release the most dependencies in the long term, but could potentially be obstructed in the short term by nodes with long execution or transmission times, thereby impeding the growth of \texttt{Nready}. Assuming the program prioritizes tasks with high data locality, or in cases where tasks with average data locality are prioritized based on higher \textit{inspiring efficiency}, \texttt{Nready} will experience rapid and significant growth.

Integrating two principles for regulating \texttt{Nready} and three methods for executing tasks, we provide detailed discussions on regulating the \texttt{Nready} indicator. The algorithm can be summarized as follows: we monitor the trend of \texttt{Nready}, and adjust the priority of corresponding tasks. During the ascending phase of \texttt{Nready}, we aim to swiftly reach the number of ready tasks that can efficiently utilize machine performance. During the stable phase of \texttt{Nready}, we seek to minimize sudden fluctuations to ensure consistent scheduling capacity for ready tasks. During the declining phase of \texttt{Nready}, we anticipate a gradual decline.

\begin{figure}[tb]
    \centering
    \begin{minipage}{0.95\linewidth}
        \caption{Pseudocode of Adaptive Priority for \texttt{Nready} Regulation.}
        \label{list:nw-code}
        \lstset{escapeinside={(*@}{@*)}}
        \begin{lstlisting}[mathescape=true,numbers=left]
init peak = 0;
if cur_nready > peak 
    peak = cur_nready;
end if
if num_nready >= peak - dec_step 
    cur_state = INC; (*@\label{state: INC}@*) 
else 
    cur_state = DEC; (*@\label{state: DEC}@*) 
end if

if cur_state == INC (*@\label{cur_state == INC}@*) 
    if cur_nready - prev_nready $\geq$ s_inc
        calculateK(cur_k); (*@\label{INC: calculateK}@*) 
        if cur_k < k_inc
            popHighEfiTask(); (*@\label{INC: popHighEfiTask}@*) 
        else if cur_k > k_inc
            popHighAbilityTask(); (*@\label{INC: popHighAbilityTask}@*) 
        end if
    end if

else if cur_state == DEC (*@\label{cur_state == DEC}@*) 
    if cur_nready > peak - s_dec $\times$ s_dec_count
        popHighAbilityTask(); (*@\label{DEC: popHighAbilityTask}@*) 
    else if cur_nready $\leq$ peak - s_dec $\times$ (s_dec_count + 1) + c
        popHighEfiLocalityTask();  (*@\label{DEC: popHighEfiTask}@*) 
        
        if cur_nready $\leq$ peak - s_dec $\times$ (s_dec_count + 1)
            updateDecrementCount(s_dec_count); (*@\label{DEC: adjustment downward}@*) 
        end if
    end if
end if
        \end{lstlisting}
    \end{minipage}
\end{figure}

Delving into the algorithm details, the corresponding pseudocode is referenced in Figure~\ref{list:nw-code}. INSPIRIT first sets up a task window to observe the trend of \texttt{Nready} and quantitatively calculates the increasing and decreasing rates in \texttt{Nready} over a period of time.
We trigger the state judgment condition when the change in \texttt{Nready} is greater than or equal to the task window. If \texttt{Nready} surpasses the previous peak, it is in the ascending state (line~\ref{state: INC}); if \texttt{Nready} falls below the previous peak, it is in the descending state (line~\ref{state: DEC}). This approach, compared to setting a time window, provides a more direct perception of the changes in the monitored parameter, enabling more timely and detailed control of \texttt{Nready}.

Next, INSPIRIT employs adaptive priority based on current states and \texttt{Nready}.
In the ascending state (line~\ref{cur_state == INC}), INSPIRIT introduces two thresholds: the ascending slope $k\_inc$ and the ascending state step size $s\_inc$, which jointly describe the expected the growth rate of \texttt{Nready}. If the growth rate of \texttt{Nready} (line~\ref{INC: calculateK}) is slow, indicating underutilization of machine performance in the past period, tasks with good data locality and high \textit{inspiring efficiency}  are prioritized to rapidly improve \texttt{Nready} (line~\ref{INC: popHighEfiTask}). Conversely, if the growth rate of \texttt{Nready} is too fast,  implying prioritization of short-term throughput in the past, the requirements for short-term machine throughput can be temporarily relaxed by considering a larger time window. In such cases, tasks with higher \textit{inspiring ability} are prioritized (line~\ref{INC: popHighAbilityTask}), ensuring that \texttt{Nready} can rapidly increase when short-term throughput needs to be improved later on.

In the stable state, if \texttt{Nready} surpasses the peak threshold, it indicates that during the preceding time interval, the machine has had sufficient ready tasks to effectively leverage its performance. Consequently, tasks demonstrating superior \textit{inspiring ability} are prioritized for execution (line~\ref{DEC: popHighAbilityTask}). Conversely, if \texttt{Nready} falls below the peak threshold, tasks offering optimal benefits in terms of both \textit{inspiring efficiency} and data locality are given precedence in execution (line~\ref{DEC: popHighEfiTask}). 
In the descending state, the difference from the ascending stage lies in the continuous decrease in task scale, necessitating a corresponding downward adjustment in the peak threshold. Therefore, we introduce a threshold, $s\_dec$, to describe the step size of the peak threshold adjustment downward. In other words, using $s\_dec$ as the granularity, the declining process of tasks is divided into multiple stages with each stage having its corresponding peak threshold. During the descending state, INSPIRIT initially behaves consistently with the stable stage. If prioritizing tasks with good data locality and high \textit{inspiring efficiency} fails to reach a sufficient number of ready tasks for machine utilization, which is the current peak threshold, the peak corresponding to the current task scale is adaptively adjusted until \texttt{Nready} returns to 0 (line~\ref{DEC: adjustment downward}).
Additionally, as INSPIRIT always observes changes in \texttt{Nready} and leverages task attributes, it can perceive runtime status and reduce the cost of adapting to different applications and hardware environments.


\subsection{Implementation details: INSPIRIT on StarPU}
\label{sec:methodology:implementation_details}
Detailed information regarding the implementation of the scheduling framework is presented in this section.
We implemented INSPIRIT on the well-established runtime system, StarPU. For the implementation of scheduling policies and scheduler modifications, StarPU incorporates \textit{push} methods for task scheduling in worker queues and \textit{pop} methods for workers to execute tasks in their respective queues based on different scheduling policies. Initially, we introduced time piling in StarPU\'s \textit{push} and \textit{pop} methods, recording the changes in \texttt{Nready}. This allowed us to observe crucial scheduling behaviors and variations in scheduling parameters. Subsequently, we calculated \textit{inspiring ability} and \textit{inspiring efficiency} using the starpu\_fxt\_tool and stored codelet performance information, incorporating them into the task attributes. Building upon this foundation, we then modified the \textit{push} and \textit{pop} methods of StarPU's built-in DMDAP policy. This modification enables adaptive task scheduling based on \texttt{Nready} through priority adjustments.

For scheduling strategy testing, Taskbench supports the analysis of various task-based runtime systems such as Legion, PaRSEC, and StarPU. However, the dependency pattern of the computational graph generated by Taskbench exhibits regularity, featuring only two topological layers of dependency, and Taskbench supports solely fully homogeneous tasks, lacking GPU implementation. To address these limitations, we followed the approach in \cite{gagrani2022neural} to generate a logically layered dataset, simulating real-scene DAG graphs. By specifying the node size of the task graph, we can generate a customized DAG storage format. Furthermore, we defined different codelets for tasks based on the inbound and outbound degree of each task node and added two implementation kernels: CPU and GPU. Subsequently, we overwrote and called the back-end interface provided by Taskbench to parse the DAG upward and access our modified StarPU downward. Specific details about DAG graph data and experimental results will be presented in Section \ref{sec:evaluation}.
\section{Evaluation}
\label{sec:evaluation}

\subsection{Experimental Setup}
\label{sec:setup}

We evaluate INSPIRIT on heterogeneous platforms with the detailed hardware and software configurations shown in Table~\ref{tab:configurations}. To validate the efficiency of INSPIRIT, we utilized StarPU, a well-established task-based runtime system, as our foundational infrastructure. Additionally, we enhance the Taskbench tool to integrate smoothly with StarPU, enabling the automated generation of task graphs. We selected automatically generated tasks as applications for evaluation in Section~\ref{sec:evaluation:evaluation_taskbench} and other real world tasks in Section~\ref{sec:evaluation:evaluation_realapp}. We test the effectiveness and scalability of INSPIRIT across a variety of applications and hardware environments, and conduct a detailed analysis of the reasons behind the effectiveness of INSPIRIT.

\begin{table}[t]
	\centering
	\footnotesize
	\caption{The hardware and software configurations.}
	\begin{tabular}{|m{40pt}<{\centering}|m{170pt}<{\centering}|} \hline
		\textbf{Platform} & {X86} \\ \hline
		\textbf{CPU} &  {Intel(R) Xeon(R) Gold 6230R CPU @ 2.10GHz} \\ \hline
		\textbf{Core} &  {26} \\ \hline
        \textbf{CUDA 0} &  {NVIDIA GeForce RTX 3090} \\ \hline
        \textbf{CUDA 1} &  {Tesla V100-PCIE-32GB} \\ \hline
		\textbf{System} &  {Ubuntu 18.04.6 LTS Linux gpu2 4.15.0-101-generic} \\ \hline
	\end{tabular}
	\label{tab:configurations}
\end{table}

\subsection{Performance on Synthesized Task DAGs}
\label{sec:evaluation:evaluation_taskbench}
\subsubsection{Taskbench}

Taskbench is a benchmark specifically designed for investigate the performance of task-based runtime systems across various conditions. It decouples the benchmark and specific runtime system implementations, and features tunable parameters such as hardware availability, task graph patterns, task granularity, and computational properties of tasks for task generation by modeling task graphs from various applications.Taskbench includes a range of common task graph patterns and representative tasks, rendering it a valuable instrument for evaluating and enhancing task-based runtime systems.


\subsubsection{Compiler Optimization}
\label{sec:evaluation_taskbench:compiler}
In order to further verify the efficiency of INSPIRIT scheduling on heterogeneous hardware environments for a variety of applications, we manually construct DAG graphs of different task sizes to verify the performance under four hardware configurations.

\textit{Implementation. }
We extended Taskbench to add logic that automates task graph generation. The algorithm of graph generation refers to the algorithm of layered graphs dataset in \cite{NEURIPS2022_6ef586bd} to simulate compiler optimization problems in the real world.

\begin{figure}[tb]
	\centering
	\includegraphics[width=\linewidth]{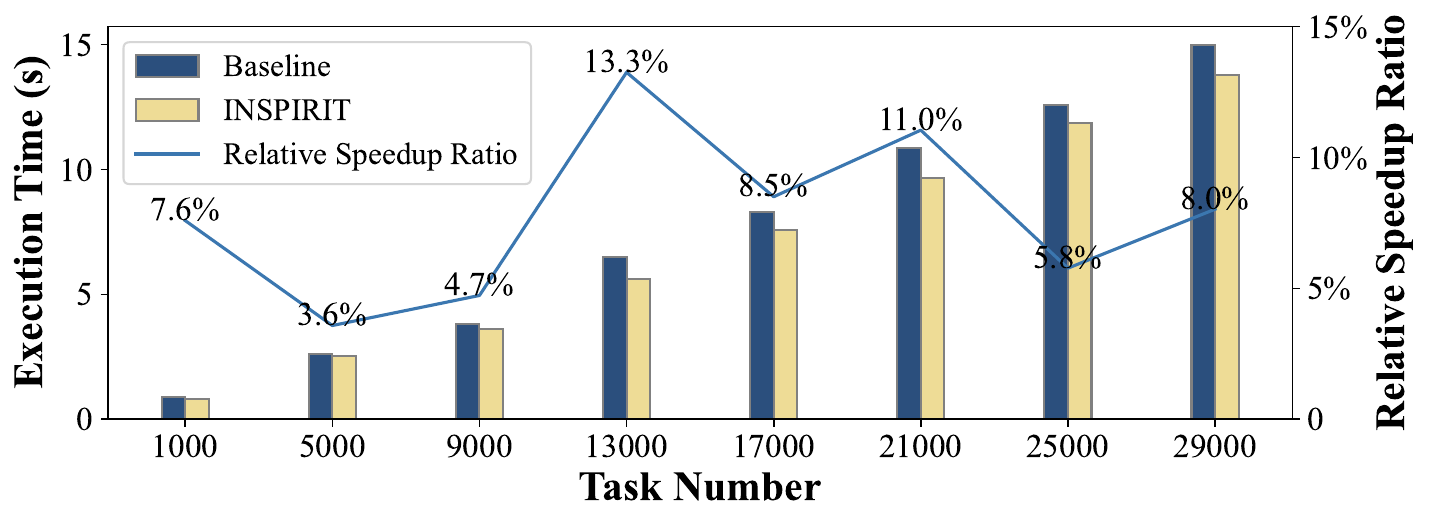}
	\caption{Effectiveness and Scalability Study on Autogen Graphs.}
	\label{fig:result_autogen}
\end{figure}

\textit{Effectiveness and Scalability. }
We tested DAGs with task sizes ranging from 1000 to 30000, in units of 4000. As shown in Figure\ref{fig:result_autogen}, we observed that INSPIRIT's performance speedup from 1.03x to 1.13x over baseline DMDA scheduling policy at different task sizes. It should be noted that since the change of task topology does not have the equivalent of mathematical task scaling, the performance improvement does not increase linearly with the change of scale.

\subsection{Performance on Real-world Task DAGs}
\label{sec:evaluation:evaluation_realapp}
\subsubsection{Cholesky}
\label{sec:evaluation_taskbench:cholesky} 
To demonstrate INSPIRIT's effectiveness, we tested Cholesky's performance on four different hardware configurations at multiple sizes. We set the block size for each task input to 960 $\times$ 960 $\times$ 4 Bytes, and block number ranging from 4 to 56.

\begin{figure*}[htbp]
	\begin{center}
	\includegraphics[width=\linewidth]{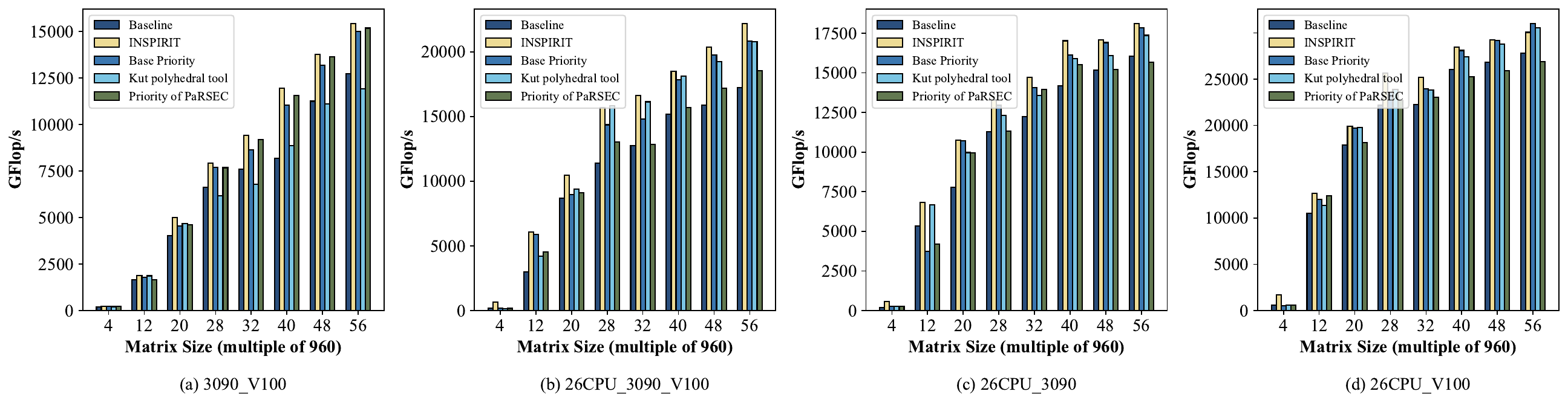}
	\end{center}
	\caption{Effectiveness Study of Cholesky Decomposition.}
	\label{fig:cholesky_effectiveness}
\end{figure*}

\textit{Effectiveness. }
As illustrated in Figure \ref{fig:cholesky_effectiveness}, INSPIRIT demonstrates performance comparable to both existing manual and automatic priority strategies across all four hardware environments. Moreover, compared to the baseline DMDA strategy, it exhibits a substantial performance enhancement, with its relative acceleration ratio varying from 1.09$\times$ to 3.22$\times$. Furthermore, INSPIRIT consistently upholds robust performance even as the size of the block Cholesky tasks increases.

\begin{figure}[tb]
	\centering
	\includegraphics[width=\linewidth]{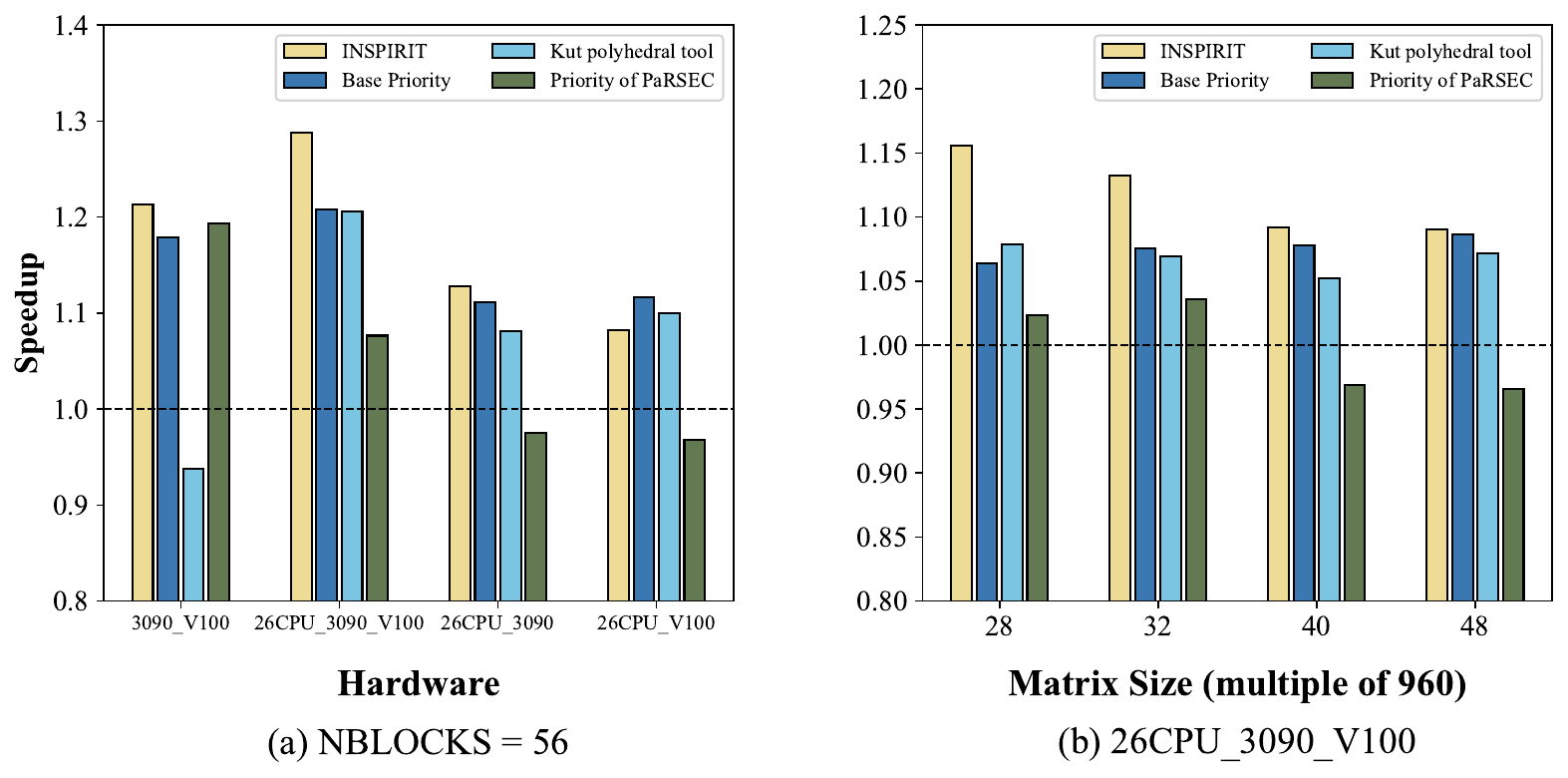}
	\caption{Scalability Study of Cholesky Decomposition.}
	\label{fig:cholesky_scalability}
\end{figure}

\textit{Scalability. } 
In order to further validate the scalability of INSPIRIT, we conduct a comprehensive analysis of experimental results from the perspectives of strategy applicability and hardware dependency.In Figure \ref{fig:cholesky_scalability}(a), we compare the speedup of various scheduling strategies against the baseline DMDA for an application with a matrix size set to NBLOCKS equaling 56. This comparison reveals that priority-based scheduling strategies based on user-configured task weights derived from domain knowledge do not consistently yield optimal performance across diverse hardware environments. In contrast, INSPIRIT, by considering the impact of hardware on execution performance, shows remarkable adaptability in various heterogeneous hardware settings.
Figure \ref{fig:cholesky_scalability}(b) maintains a constant hardware setup of 26CPU\_3090\_V100 and examines the speedup of different scheduling strategies compared with the baseline DMDA. It is clear that priority-based scheduling strategies based on user-configured task weights derived from domain knowledge encounter difficulties in preserving scalability of performance. INSPIRIT, however, by acknowledging the strong link between execution performance and application characteristics, consistently exhibits stable and improved relative speedup across a wide spectrum of applications and different task scales within the same application.

\textit{Analysis. } 
To illuminate the origins of performance enhancement in INSPIRIT, we instrumented an application with a fixed block number of 64 and utilized a 26CPU\_3090 hardware environment. This setup was employed to chronicle the temporal progression of key intermediate variables in both the state-of-the-art (SOTA) scheduling strategy and the INSPIRIT scheduling process. Specifically, our focus was on tracking the fluctuations in Nready, push, and pop actions over time. We have documented and elucidated two distinct categories of figures: the Nready Trend plots (Nready\_time plots) and the Schedule Operations plots (push\_pop plots).

\begin{figure}[tb]
	\centering
	\includegraphics[width=0.8\linewidth]{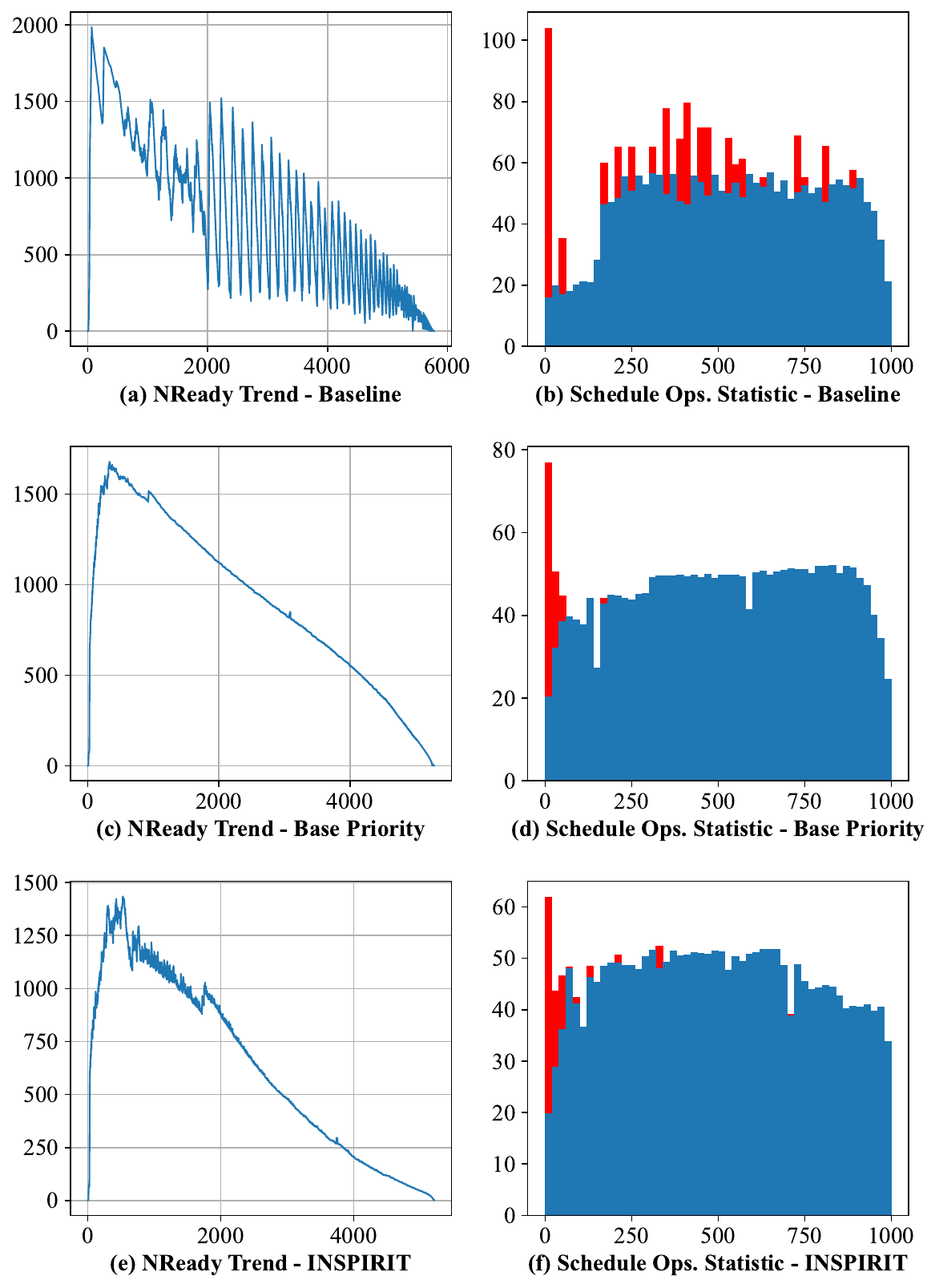}
	\caption{Analysis on Cholesky Decomposition.}
	\label{fig:analysis_on_cholesky_decomposition}
\end{figure}

The Nready\_time plots serve to illustrate whether the machine's workers have a sufficient number of tasks to execute, reflecting the changing trend in the number of schedulable tasks in the queue over time. All three scheduling strategies initially ensure that the number of tasks observed during execution exceeds or equals the number of available workers. Subsequently, we examine the push\_pop plots. For the push\_pop plots, we focus on two sets of data: the number of schedulable tasks generated within a specific timeframe (indicated by red bars, triggered by push actions), and the number of tasks executed within the same timeframe (represented by blue bars, triggered by pop actions). When the red bars are higher than the blue bars, it indicates that task execution is constrained by machine performance or data transfer during that period. Conversely, when the blue bars exceed the red bars, it suggests that task execution is constrained by the availability of schedulable tasks within the machine, with the possibility of achieving peak performance.
Figure~\ref{fig:analysis_on_cholesky_decomposition} presents, from left to right, the Nready\_time and push\_pop plots for baseline DMDA, the SOTA strategy (Base Priority), and INSPIRIT. 

An initial observation from the Nready\_time plots reveals that all three scheduling strategies initially ensure a surplus of schedulable tasks, with the number of tasks observed consistently exceeding the number of available workers.
Further examination of the push\_pop plots reveals distinctive characteristics. The baseline DMDA strategy exhibits a significant horizontal execution of SYRK tasks in the early stages (as seen in Figure~\ref{fig:Cholesky_dag}), which inadequately inspire subsequent tasks. Consequently, during a substantial duration, as shown in Figure~\ref{fig:analysis_on_cholesky_decomposition}(d), the number of popped tasks cannot even reach 25. 
A similar, though less pronounced, issue is observed for the SOTA strategy between 0ms and 250ms, indicating a performance limitation due to insufficient task inspiration. Upon comparison of the SOTA strategy with INSPIRIT, it becomes apparent that both strategies experience performance limitations due to insufficient task inspiration within the normalized time interval of 500ms to 750ms. However, the SOTA strategy encounters this issue between 0ms and 250ms as well. Therefore, INSPIRIT exhibits a slight advantage over the SOTA strategy in this regard.

Finally, revisiting the temporal evolution of Nready, we observe distinct patterns. Under the baseline DMDA strategy, Nready exhibits sudden increases and decreases over time, indicating abrupt changes in the number of schedulable tasks. Such behavior implies that the number of schedulable tasks consistently experiences sudden increments and decrements, preventing the machine's performance from being fully utilized and leading to resource wastage.
Conversely, the Nready curve for the SOTA strategy displays a smoother temporal evolution. This approach, based on human expertise, objectively manages task growth and reduction, ensuring a seamless transition. However, it suffers from inherent drawbacks such as prolonged development time, high domain knowledge requirements, and lack of sustainability. Additionally, it is highly application-dependent. The performance enhancement observed in INSPIRIT, in comparison to such strategies, further underscores the potential benefits of INSPIRIT, implying that it might not exhaust optimization opportunities. INSPIRIT, guided by observations of Nready, adaptively adjusts priorities to govern task execution. In doing so, it avoids abrupt curve transitions observed in the Nready plot while indirectly showcasing its scalability.

\subsubsection{LU}
\label{sec:evaluation_taskbench:lu}
To demonstrate INSPIRIT's effectiveness, we tested LU's performance on four different hardware configurations at multiple sizes. We set the block size for each task input to 160 $\times$ 160 $\times$ 4 Bytes, and block number ranging from 36 to 52.

\begin{figure*}[htbp]
	\begin{center}
	\includegraphics[width=\linewidth]{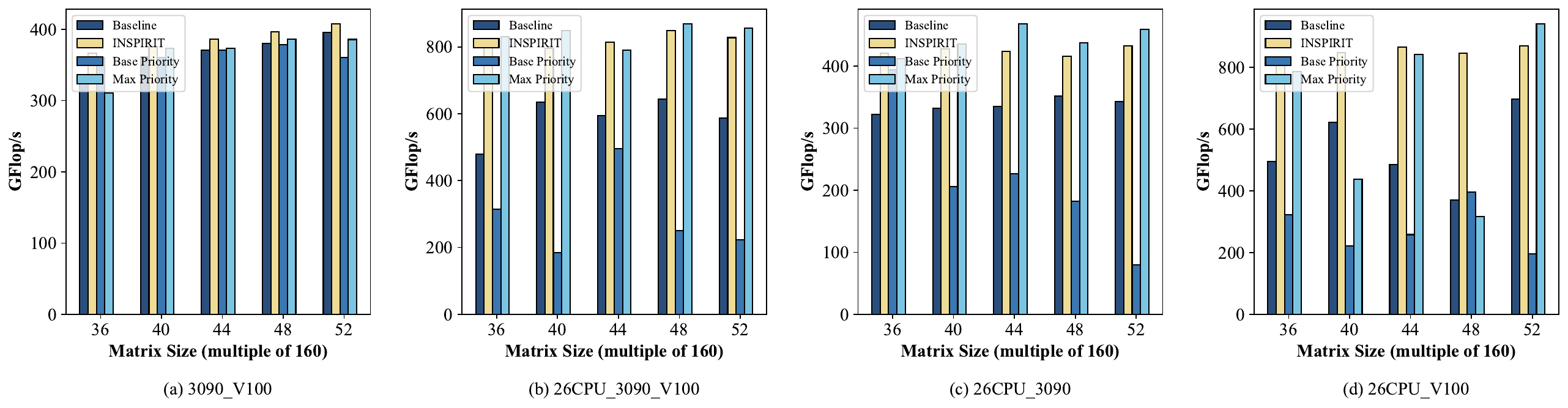}
	\end{center}
	\caption{Effectiveness Study of LU Decomposition.}
	\label{fig:lu_effectiveness}
\end{figure*}

\textit{Effectiveness. }
From Figure~\ref{fig:lu_effectiveness}, it can be observed that, across four different hardware environments, INSPIRIT demonstrates performance comparable to existing manual and automatic priority-based strategies. Furthermore, it exhibits a notable performance improvement relative to the baseline DMDA strategy, with a range of relative speedup values spanning from 1.03$\times$ to 2.28$\times$. 
\begin{figure*}[htbp]
	\begin{center}
	\includegraphics[width=\linewidth]{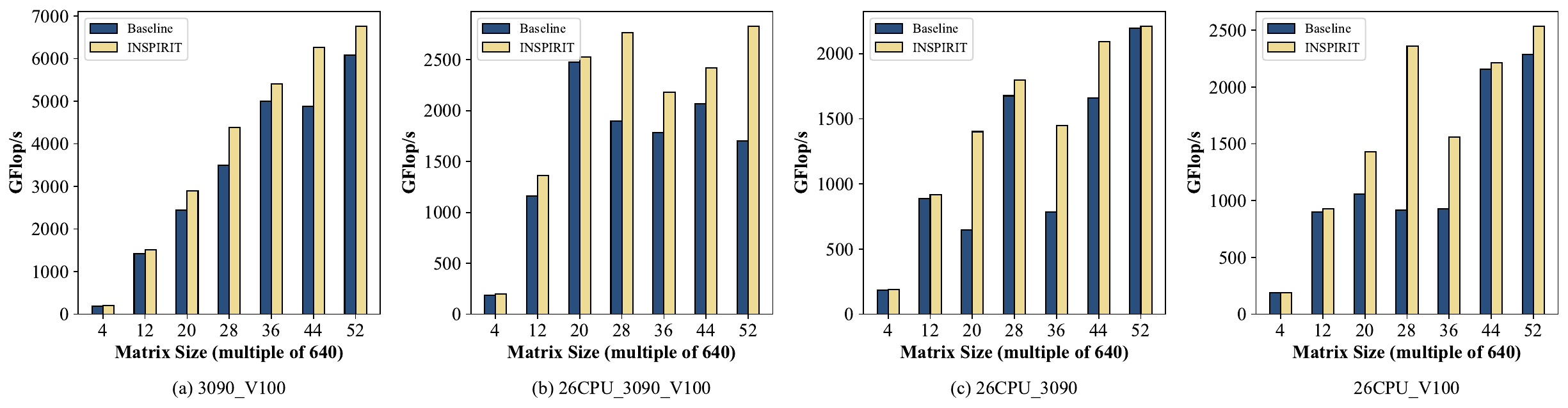}
	\end{center}
	\caption{Effectiveness Study of Heat.}
	\label{fig:heat_effectiveness}
\end{figure*}

\begin{figure}[tb]
	\centering
	\includegraphics[width=\linewidth]{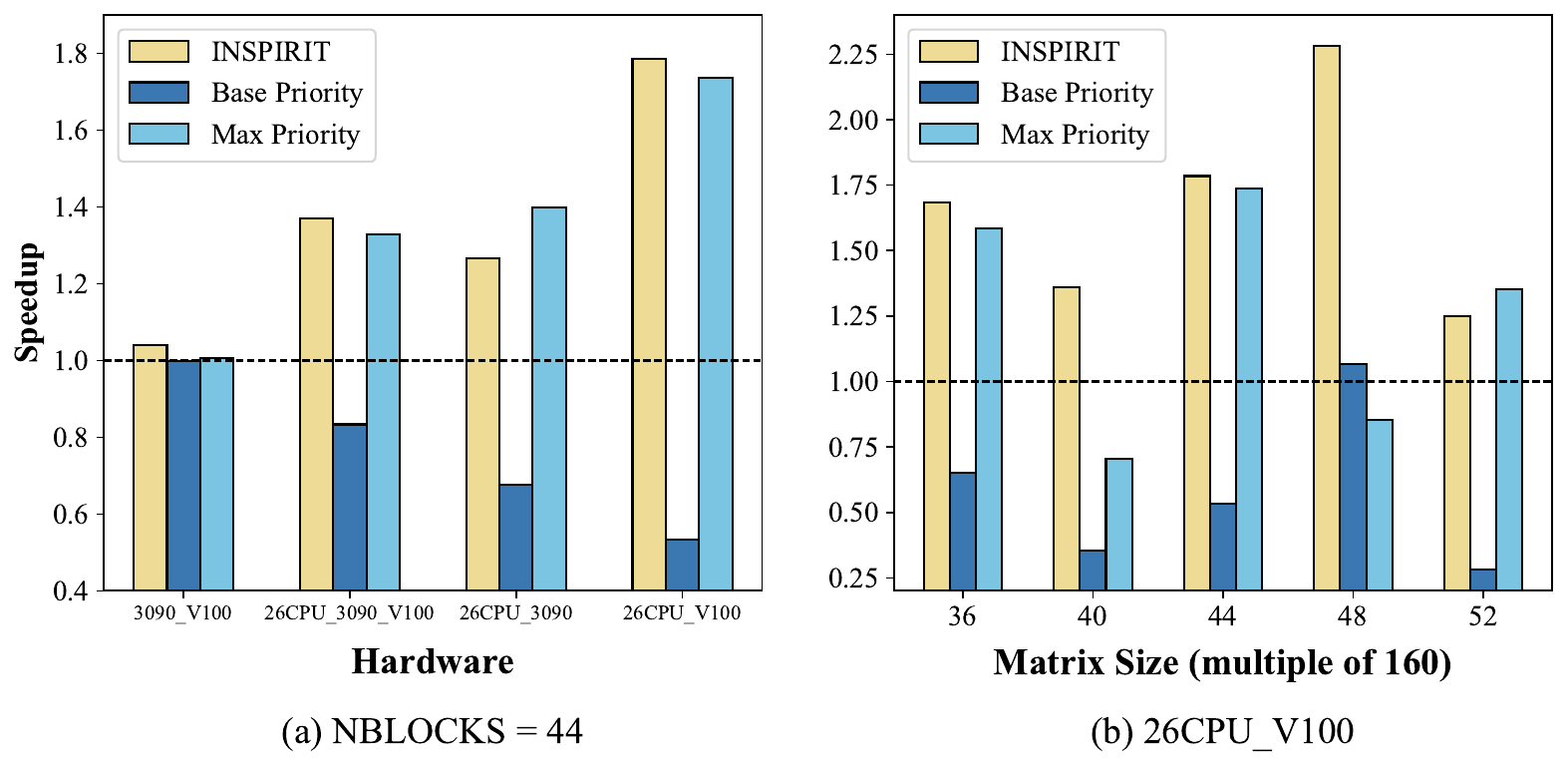}
	\caption{Scalability Study of LU Decomposition.}
	\label{fig:lu_scalability}
\end{figure}

\textit{Scalability. } 
We further corroborated the scalability of INSPIRIT in the context of LU decomposition tasks, building upon the aforementioned experimental results. Figure \ref{fig:lu_scalability}(a) displays the relative scalability ratios of various scheduling strategies compared to the baseline DMDA, specifically for LU decomposition with a matrix size set to NBLOCKS of 44. This figure highlights that scheduling strategies relying on priorities based on user-configured task weights derived from domain knowledge do not consistently achieve optimal performance across diverse hardware environments. INSPIRIT, by accounting for the impact of hardware on execution performance, demonstrates adaptability to a range of heterogeneous hardware contexts.
In Figure \ref{fig:lu_scalability}(b), the relative scalability of DMDA under different scheduling policies is showcased, with the hardware configuration fixed at 26CPU\_V100. It is evident from this figure that scheduling policies based on manual optimization priorities struggle to maintain scalability. In contrast, INSPIRIT recognizes the strong correlation between execution performance and application characteristics, achieving stable and favorable relative acceleration ratios in various applications and even across different task scales within the same application.


\subsubsection{Heat}
\label{sec:evaluation_realapp:heat}
In order to prove the scheduling effectiveness and scalability of INSPIRIT in real applications, we select a real case of solving heat transfer problems. The heat program is a parallel numerical solver for heat transfer problems\cite{Fourier2009ThorieAD}, employing the finite element method to solve PDEs across a mesh. The mesh size is directly related to the accuracy and time consumption of the simulation, and is adjustable according to the specifics of the problem at hand. Thus, we choose a block size of 640 $\times$ 640 $\times$ 4B and then verify the scheduling performance of INSPIRIT under four hardware configurations on block numbers range from 4 to 52.

\textit{Effectiveness and Scalability. }
Figure \ref{fig:heat_effectiveness} demonstrates that INSPIRIT is capable of efficiently solving heat tasks of varying scales across four distinct hardware environments. When compared to the baseline DMDA scheduling policy, INSPIRIT achieved an end-to-end performance improvement of up to 1.58$\times$.
\section{Related Work}
\label{sec:relatedwork}

Scheduling DAGs is a classical NP-Complete (NPC) problem, inherently unsolvable in polynomial time in its general case. In general, DAG scheduling techniques can be categorized into static and dynamic DAG scheduling. In static scheduling, all schedule decision is made before execution. Conversely, dynamic scheduling involves making decisions during execution, based on real-time metrics as well as pre-existing information. For static DAG scheduling, prevalent strategies encompass list scheduling, optimization algorithms, and learning-based approaches. 

\textbf{List scheduling - }List scheduling assigns tasks in a list to processors by predefined criteria. This type of scheduling includes earlier DAG scheduling efforts such as DCP~\cite{kwok1996dynamic}, MCP~\cite{kwok1996dynamic}, HEFT and CPOP~\cite{topcuoglu2002performance}, SDLS~\cite{li2013scheduling}. The DCP algorithm optimizes schedule length by computing critical paths and minimizing them at each scheduling step. The HEFT algorithm analyzes the critical path's length from the current task's start to the exit task, considering the earliest start time from the entry task. The CPOP algorithm further computes critical path for each processor and takes communicate cost into account. While the core idea of utilizing idle slots during data transmission is realized in these algorithms, the complexity and time consumption of this analysis limit their scalability for larger tasks. Other critical path-based scheduling algorithms encounter similar scalability issues. These classic algorithms all focus on minimizing execution makespan and prioritizing tasks accordingly. Recently, attention has shifted towards optimizing other objectives such as energy efficiency. Notable examples include REAS and RHEFT~\cite{5376540}.

\textbf{Optimization algorithms - }Furthermore, it is often imperative to schedule DAGs with multi-objective considerations, leading to the adoption of metaheuristic optimization methods, such as the Simulated Annealing or Genetic Algorithm. Nevertheless, the expansive search space associated with these algorithms frequently results in convergence times spanning hours or days. Numerous studies have been undertaken to mitigate this challenge, illustrative works including Constraint Programming~\cite{constrainBasedScheduling}, NGA~\cite{driss2015new} and AutoMap~\cite{SFXTeixeira2023AutomatedMO}. NGA accelerates scheduling by employing a semi-conducted population in the genetic algorithm, where certain individuals are generated using the HEFT algorithm and others are generated randomly. AutoMap focuses on optimizing the execution of tasks while balancing the trade-off of minimizing data movement, thereby curtailing the search space. However, the DAGs frequently need to be rescheduled in response to changes in applications or hardware, and the iterative profiling required for these methods remains a time-intensive endeavor.

\textbf{Learning based scheduling - }Learning based scheduling mainly includes GNN. As DAG belongs to a kind of graph, it is a direct idea to use graph neural network (GNN) to solve DAG scheduling problem. DAG graph can be directly connected to GNN~\cite{thost2021directed, zhang2019d} simply by adding edge information. However, on the one hand, the goal of generating scheduling order is too high for the model to generate feasible solutions accurate to the processor; on the other hand, the differentiated node features learned by GNN are mainly derived from the neighbor distribution, which conflicts with the scheduling of information that needs to be global.


In the realm of dynamic DAG scheduling, existing static scheduling approaches, notably list scheduling, are frequently integrated with historical execution data to develop dynamic scheduling methodologies. One such example is the EET~\cite{2013An} policy. This dynamic approach involves the scheduler recording runtime data, such as kernel execution times, to inform subsequent scheduling decisions. Similarly, StarPU's DMDAP policy represents another dynamic variant grounded in list scheduling principles. The approach presented in this paper aligns with these dynamic scheduling strategies.
The scheduling algorithm based on reinforcement learning and reinforcement learning (RL)~\cite{wu2018adaptive} is also a typical dynamic scheduling policy and are often designed to optimize task allocation among processors. However, deep reinforcement learning approaches in scheduling face challenges including the need for extensive datasets, complex modeling, prolonged training durations, and significant resource consumption for training.
Additionally, there are dynamic scheduling strategies that defy straightforward categorization, such as work stealing~\cite{blumofe1999scheduling, workStealing2}, which ensures load balancing by enabling idle resources to autonomously acquire tasks from busier resources. It is important to note that these strategies like work stealing operates orthogonally to the methodologies discussed in this paper.

Unlike the aforementioned scheduling policies, INSPIRIT leverages the full spectrum of heterogeneous hardware capabilities by abstracting task attributes. It dynamically adapts to variations in upper-layer applications through interfaces provided by the Task-based runtime system, consistently delivering stable performance. In INSPIRIT, the \textit{inspiring ability} of all tasks can be determined at the cost of a single traversal, a negligible process as most scheduling algorithms necessitate such traversal for global information. Moreover, the computation of \textit{inspiring efficiency} employs a slicing window approach, limiting time consumption, ensuring that the cost of computing priority remains manageable. Furthermore, INSPIRIT utilizes the count of executable tasks in all worker queues as a metric to direct task execution. By merely monitoring their fluctuations, INSPIRIT efficiently minimizes the overhead associated with aligning tasks to diverse hardware resources.

\section{Conclusion}
\label{sec:conclusion}
In this paper, we introduce INSPIRIT, an efficient and lightweight DAG scheduling framework tailored for task-based runtime systems. INSPIRIT can efficiently adapt to various applications and achieve better performance across different hardware platforms while maintaining low overhead and obviating the need for user awareness regarding specific applications or hardware configurations. At the same time, INSPIRIT provides better extensibility by only using abstracted collectively by existing task-based runtime systems. Specifically, we define \textit{inspiring ability} and \textit{inspiring efficiency} as task attributes to regulate changes in \textit{Nready} during execution. Meanwhile, INSPIRIT monitors \textit{Nready} to guide task scheduling sequence. By evaluating INSPIRIT on both real-world DAGs and auto-generated DAGs based on StarPU, we demonstrate that INSPIRIT can effectively schedule various tasks on heterogenous platforms with performance speedup ranging from 1.03$\times$ $\sim$ 3.22$\times$ compared to state-of-the-art static and dynamic scheduling polices.

%

%
\bibliographystyle{ACM-Reference-Format}
\bibliography{ref.bib}

%
\end{document}